\documentstyle[epsfig]{aipproc}
\begin{document}
\title{Distribution of Fluences versus Peak Fluxes  and the
Cosmological Density Evolution of the Rate of GRBs}

\author{ Vah\'e Petrosian and  Nicole M. Lloyd }
\address{Center for Space Science and Astrophysics \\ Stanford University\\
Stanford, California 94305}

\maketitle

\begin{abstract}

We argue that recent observations of afterglows and some theoretical considerations indicate that
the gamma-ray fluence may be a better measure of strength of a burst than its peak 
flux.  We discuss how the distribution of the fluence, or any physical quantity related to the peak counts (or
 flux), can be obtained with proper correction for threshhold effects.  Using this method, we
 compare the cumulative and differential distributions of flux and fluence.
 We find some differences between these distributions, but more remarkable are the similarities
 between these distributions.  The other striking feature is how different these 
 distributions are from those of other
 extragalactic objects.  Using the fluence distributions, we derive the expected comoving density rate
 evolution for different assumed total luminosity; we compare this with the 
GRB rate expected from the 
 recently determined
 star formation rate.  
\end{abstract}

\section*{Introduction}

In the absence of knowledge of redshifts or direct distances to a large
number of gamma-ray bursts (GRBs), we have to rely on the distributions
of important physical parameters such as flux, fluence, duration and 
spectral characteristics, as well as the correlations between these parameters,
for further
insight into the physics of the energization and radiation of these sources.
One approach is to start with specific parametrized models and compare
the expected results with observations. Inclusion of all observational 
constraints makes such models very complex. An alternative approach
(the one we prefer until more redshifts are known) is to work
from the observations to obtain bias free, 
non-parametric distributions. 

As the name ``gamma ray burst" implies, the peak gamma-ray flux $f_{p}$ of a GRB far exceeds the
peak fluxes at other energies. The recent X-ray, optical and radio observations 
of the afterglows - in particular the fact that these fluxes decline
more rapidly than $1/t$ (see e.g. Van Paradijis in these proceedings) - 
indicate that the energy fluence $F$ in the gamma-ray
range is also higher than in any other band. Therefore, the gamma-ray
fluence $F$ provides the best measure of the strength of the burst.
This is also what is expected theoretically from the fireball model, where the 
total radiant luminosity is expected to be a standard candle, representing
the released graviational potential, whereas the peak luminosity, duration,
etc. (which are affected by the variable values of the bulk Lorentz factor or
baryon loading of the fireball) are not expected to be standard measures.

\begin{figure}
\centerline{ \epsfig{file=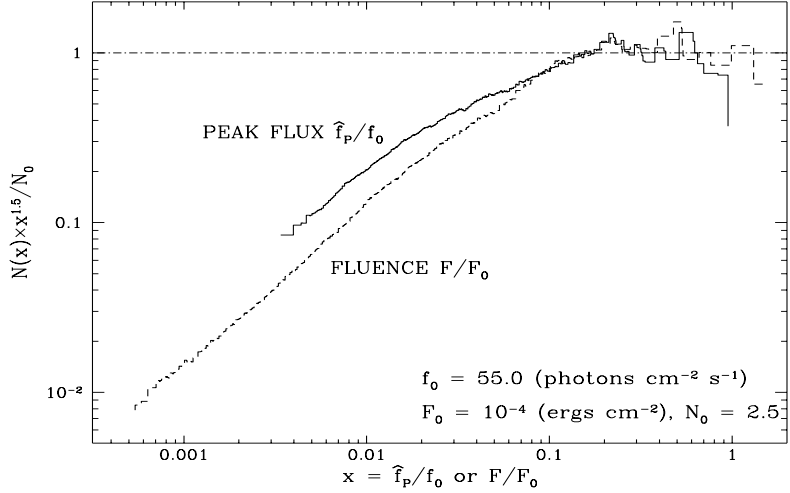, width=11cm,height = 7cm,angle=270}}
\caption{The cumulative counts multiplied by the 3/2 power of the variable
to remove the Euclidean (more correctly HISE) dependence for fluxes (solid
histogram) and fluences (dashed histogram) of the 3B GRBs.}
\end{figure}

In spite of this, most analysis such as the Log$N$-Log$S$, time dilation, etc.
have been carried out using the peak photon flux with the tacit assumption that
the peak photon luminosity is a representative measure of the burst strength.
The primary reason for this is because the burst triggering is based on
photon counts (or flux) and not on the fluence. However,
this should not deter us from using the fluences because,
as argued by Petrosian and Lee \cite{PetLee}, the existing data can be used 
to determine the limits on the fluences of the bursts. 
In fact, this argument can be generalized
to any physical quantity $X$ related to the parameter which determines
the triggering threshold, which is, in this case, the peak flux $f_p$ or the peak count
$C_{\rm {max}}$. For example, given $f_p$, its threshold $f_{lim}$ and 
the relation
$X(f_p)$, then in the spirit of the well known $V/V_{max}$ test, we can ask:
What is the range of possible values of $X$ so that the peak flux exceeds
the threshold? The limiting values $X_{\rm {min}}$ and/or $X_{\rm {max}}$ are obtained
from $X(f_{lim})$.

Quantities related to the flux monotonically, e.g. fluence, will have only
a lower limit while those with a more complex relation may have both
an upper and a lower bound. In our past works in this area we have 
developed and utilized non-parametric methods to determine the bias free
distributions and correlations from data truncated on one side \cite{VP,EP92}.
We have recently generalised these to the two sided truncation case \cite{New}.
An example of such a case is discussed by us elsewhere in these 
proceedings \cite{NicP97},
dealing with the distribution of the peak energy $E_p$ of thr $\nu F(\nu)$
spectrum. Here we use the one sided methods and compare the distributions
of the flux and fluence. 

\begin{figure}
\centerline{\epsfig{file=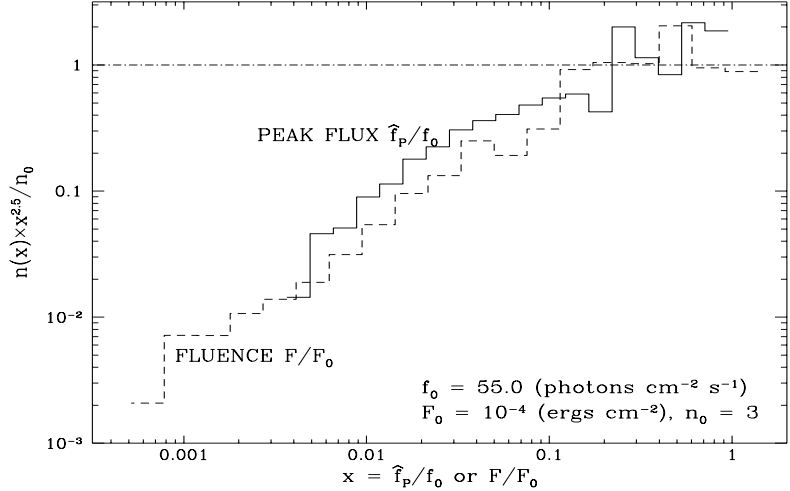, width = 11cm,height=7cm,angle=270}}
\caption{The differential counts multiplied by the 5/2 power of the variable
to remove the Euclidean (more correctly HISE) dependence for fluxes (solid
histogram) and fluences (dashed histogram) of the 3B GRBs.}
\end{figure}

\section*{Distributions of Flux and Fluence}
 
Figure 1 shows the cumulative distributions of peak fluxes and fluences of the
GRBs in the 3B catalog, where the bias at low values of these quantities 
due to the threshold and the effects of correlations have been corrected for. For details
the reader is refered to our earlier papers \cite{PetLee,VP,LeeP96}. 
We plot the deviations
from the -3/2 power law dependence expected in the homogeneous, isotropic,
static and Euclidean (HISE) model.

For both quantities there are well defined 3/2 power law portions
and a sharp deviation from this, especially for the fluence.
This may be an indication that the total luminosity has a narrower
distribution (is a better standard candle) than the peak luminosity.
This and the larger range of the fluence makes it a more appropriate 
parameter for cosmological tests.

The differential distributions divided by that expected for HISE case are
shown in Figure 2. The above mentioned difference is evident here, although
to a lesser degree.
A more striking feature is the remarkable
similarity between the shapes of the two distributions. 
These differences and similarities can be also seen in Figure 3. 

\begin{figure}
\centerline{\epsfig{file=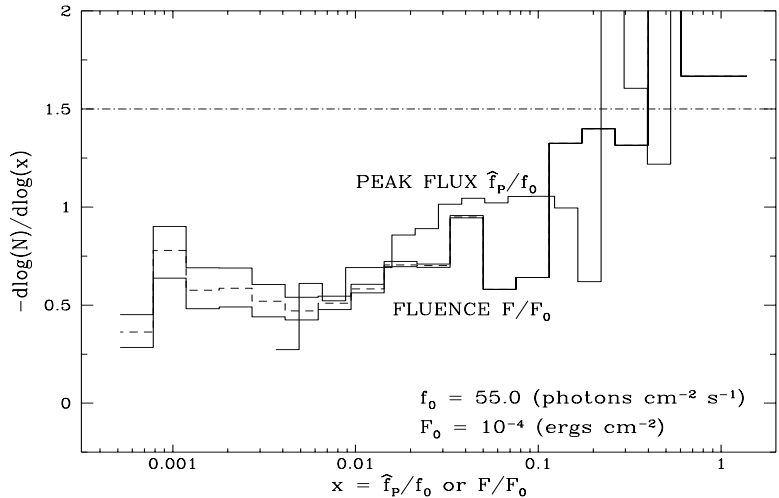, width = 11cm,height=7cm,angle=270}}
\caption{The logarithmic slope of the cumulative counts for fluxes (solid
histogram) and fluences (dashed histogram) of the 3B GRBs. The thin 
solid lines show the 90\% confidence limits on the corrections used for the 
threshold and correlation effects (1)}
\end{figure}

\section*{Rate Evolution and Discussion}

The above counts are unlike the counts of other well known extragalactic
sources such as galaxies, radio sources and AGNs or quasars. The transition
from 3/2 power law is too abrupt and the slope beyond this transition is
nearly constant, especially for the fluence. Clearly, some extraordinary evolutionary
processes are at work.  There has been several detailed analyses of the distributions
of the fluxes (see e.g. references \cite{LorWass} or \cite{Hakkila}) with inconclusive
results.  This is primarily because 
any observed distribution can be fitted to an arbitrary luminosity function and evolution even if one
assumes a cosmological model.  To obtain
some indication of possible evolutions,  we assume several representative values for the emitted energy or the total
luminosities $\cal L$ and derive the comoving rate density
from observed differential counts of the fluences; 
\begin{equation}\label{equation}  
        \rho(z) =(1 + z)n(F)(dz/dV)(dF/dz).
\end{equation}
Here $F = \cal L$ $/(4\pi d_L^{2} (1 + z)^{\beta - 3})$; 
$d_{L} = (2c/H_0)(1 + z - (1 + z )^{1/2})$
is the luminosity distance in an $\Omega = 1, \Lambda = 0$ cosomological
model, and $-\beta$ is the photon flux spectral index.  
Figure 4 shows this rate evolution for three representative 
values of $\cal L$ and for two
spectral indices.
\begin{figure}
\centerline{\epsfig{file=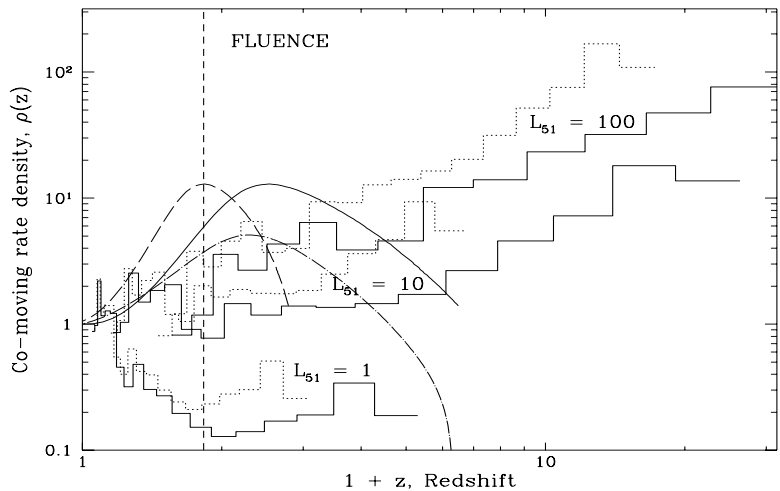, width = 11cm,height=7cm,angle=270}}
\caption{Rate per unit co-moving volume of GRBs as a function of redshift
obtained from the differential distribution of the fluences for three different
energies of GRBs ($L_{51} = $ $\cal L$ $/(10^{51}$ ergs) and two different 
values of the photon spectral index: $\beta = 2$ (solid histogram), 
$\beta = 3$ (dotted histogram). The vertical dashed line shows the 
location of the redshift of the May 8, 1997 afterglow. 
The continuous curves show three representative rates expected from
the star formation rate (see text). $H_{o} = 60 {\rm km s^{-1} Mpc^{-1}}$.} 
\end{figure}
  For a low energy source $\cal L$ $= 10^{51}$ erg/s, the density decreases with redshift, while
  it increases for high energies. The observed redshift of 0.83 for GRB 970508
  indicates that the most likely energy is about $\cal L$ $= 10^{52}$ erg/s with possibly little evolution.  
    
    Figure 4 also shows the star formation rate from Madau (\cite{Madau}, 
solid curve), this rate 
    delayed by $2\times 10^{9}$ years (dashed curve), and the rate convolved with a distribution
    of delays $P(t) \propto t^{-1}$ (dot-dashed curve; see ref. 
\cite{Tot}).
 As evident, the observed evolutions are not similiar
    to that expected from the star formation rate, indicating a complex distribution
    of $\cal L$ and the presence of other evolutionary processes.

\end{document}